\documentclass[pra,superscriptaddress,twocolumn,superscriptaddress,amsmath]{revtex4}

\usepackage{amsmath, amsthm, amssymb}
\usepackage[]{graphicx} 
\usepackage{dcolumn} 
\usepackage{bm} 
\usepackage[normalem]{ulem}
\usepackage{cases}
\usepackage{color}
\usepackage[dvipsnames]{xcolor}
\usepackage{algorithm}
\usepackage{algpseudocode}
\usepackage{csquotes}
\usepackage{mathtools}
\usepackage{mathrsfs}
\usepackage[caption=false]{subfig}
\usepackage{lipsum}
\usepackage{comment}
\usepackage{tabularx}
\usepackage{multirow}
\usepackage{braket}
\usepackage{pgf} 
\usepackage{siunitx} 

\usepackage{varioref}  
\usepackage{hyperref}
\hypersetup{
    colorlinks=true,
    linkcolor=blue,
    citecolor=blue,
    filecolor=magenta,      
    urlcolor=cyan
    }
\usepackage{cleveref}   

\usepackage[colorinlistoftodos]{todonotes}

\definecolor{pinocol}{rgb}{0,.6,.2}

\begin{document}

\title{Long-term analysis of \textit{efficient-BB84} 4-node network with optical switches in metropolitan environment}

\author{Alberto De Toni}
\affiliation{Dipartimento di Ingegneria dell'Informazione, Università degli Studi di Padova, via Gradenigo 6B, IT-35131 Padova, Italy}

\author{Edoardo Bortolozzo}
\affiliation{Dipartimento di Ingegneria dell'Informazione, Università degli Studi di Padova, via Gradenigo 6B, IT-35131 Padova, Italy}
\affiliation{ThinkQuantum s.r.l., via della Tecnica 85, IT-36030 Sarcedo, Italy}

\author{Alessandro Emanuele}
\affiliation{ThinkQuantum s.r.l., via della Tecnica 85, IT-36030 Sarcedo, Italy}

\author{Marco Venturini}
\affiliation{ThinkQuantum s.r.l., via della Tecnica 85, IT-36030 Sarcedo, Italy}

\author{Luca Calderaro}
\affiliation{ThinkQuantum s.r.l., via della Tecnica 85, IT-36030 Sarcedo, Italy}

\author{Marco Avesani}
\affiliation{Dipartimento di Ingegneria dell'Informazione, Università degli Studi di Padova, via Gradenigo 6B, IT-35131 Padova, Italy}
\affiliation{ThinkQuantum s.r.l., via della Tecnica 85, IT-36030 Sarcedo, Italy}

\author{Giuseppe Vallone}
\affiliation{Dipartimento di Ingegneria dell'Informazione, Università degli Studi di Padova, via Gradenigo 6B, IT-35131 Padova, Italy}
\affiliation{ThinkQuantum s.r.l., via della Tecnica 85, IT-36030 Sarcedo, Italy}
\affiliation{Padua Quantum Technologies Research Center, Università degli Studi di Padova, via Gradenigo 6A, IT-35131 Padova, Italy}

\author{Paolo Villoresi}
\affiliation{Dipartimento di Ingegneria dell'Informazione, Università degli Studi di Padova, via Gradenigo 6B, IT-35131 Padova, Italy}
\affiliation{ThinkQuantum s.r.l., via della Tecnica 85, IT-36030 Sarcedo, Italy}
\affiliation{Padua Quantum Technologies Research Center, Università degli Studi di Padova, via Gradenigo 6A, IT-35131 Padova, Italy}

\date{\today}

\begin{abstract}

Quantum Key Distribution (QKD) is a leading technology for enabling information-theoretic secure communication, with protocols such as BB84 and its variants already deployed in practical field implementations. As QKD evolves from point-to-point links to multi-node networks, scalability and cost-effectiveness become central challenges. Among the approaches to address these issues, efficient-BB84 has shown durable and reliable performances, while optical switching techniques enable flexible, scalable, and cost-efficient integration of QKD into existing infrastructures. In this work, we present an active QKD network in a production environment, employing efficient-BB84 and optical switching, orchestrated in a coordinated manner, emphasizing their potential to support robust, future-proof quantum-secure communication systems.

\end{abstract}

\maketitle

\section{Introduction}
Quantum Key Distribution(QKD) represents one of the most mature applications of quantum information science, offering unconditional security for key exchange based on the fundamental laws of quantum mechanics \cite{bennettQuantumCryptographyPublic2014, Gisin2002}. Protocols such as BB84 and decoy-state variants have been implemented over increasing distances and integrated into field-deployable systems \cite{lucamariniOvercomingRateDistance2018, Wang2022}.

QKD is emerging as a cornerstone technology for achieving information-theoretic secure communication, leveraging the fundamental principles of quantum mechanics~\cite{bennettQuantumCryptographyPublic2014, gisinQuantumCryptography2002}. While point-to-point QKD links have demonstrated secure key exchange over both fiber and free-space optical channels, the integration of QKD into scalable, multi-node networks is essential for widespread adoption in real-world applications, such as secure governmental communication, financial services, and critical infrastructure protection~\cite{chen2021integrated}.

QKD networks introduce new challenges and design requirements, including efficient key management, routing, and node interoperability.
Recent network implementations have adopted efficient-BB84 to improve scalability and performance, enabling more practical deployments across metropolitan and backbone-scale infrastructures~\cite{Peev_2009, Dynes2019}. These advancements make efficient-BB84 a promising candidate for future quantum-secure communication networks, particularly in scenarios where channel loss and system complexity must be tightly controlled.
Among the other various techniques employed in QKD networks, physical link switching has gained particular attention due to the reduced cost of technology introduction and the possibility of iterative scalability.
By changing the wavelength or the lightpath of qubits, different implementations allow for more flexibility both in infrastructure design and on-demand performance. 

In this paper, we present and analyze VenQCI, a production network designed and realized by the Veneto Region, Concessioni Autostradali Venete (CAV) and University of Padova,
the first regional quantum network in Italy, and part of the future Italian quantum communication infrastructure.
This network was planned and implemented with industrial usage as the final objective, with ease of maintainability and minimal future expansion cost as key points, leveraging efficient BB84 and link switching.
We obtained a stable network that, as of the time of writing, is used for the encryption of traffic data, like highway webcam video streams, and is suitable to support other security applications.

\section{Description of the devices}
In this section we will describe the QKD devices and the network orchestration software employed in the present work.

\subsection{QuKy by ThinkQuantum s.r.l.}

\begin{figure}[h]
    \centering
    \includegraphics[width=0.8\linewidth]{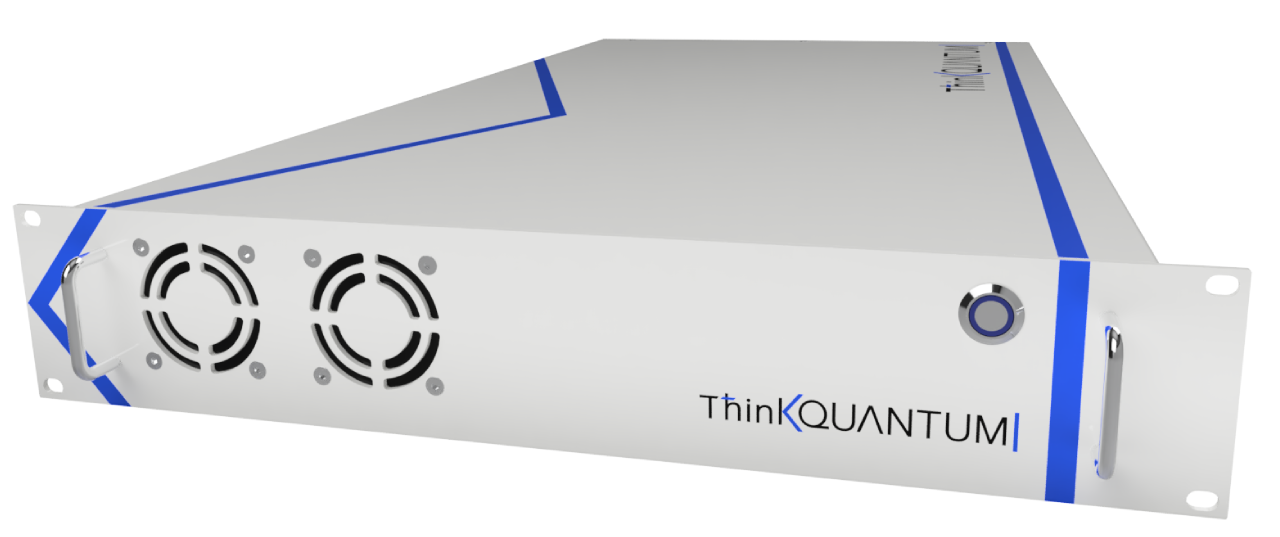}
    \caption{QuKy by ThinkQuantum s.r.l. \cite{thinkquantum}}
    \label{fig:QuKy}
\end{figure}

QuKy~\cite{thinkquantum} (see \vref{fig:QuKy}) is a ThinkQuantum QKD platform implementing BB84 for secure key distribution. It integrates a QRNG delivering a direct qubit stream, eliminating key expansion and pseudorandomness, thus enabling high key rates with strong security and robustness.

\subparagraph{Compatibility and Flexibility}
QuKy supports multiple network topologies and key-management protocols, operating over both fiber and free-space optical links for broad deployment options.

\subparagraph{Interoperability}
Alice and Bob require no dedicated pairing: a single unit can interoperate with multiple counterparts, enhancing versatility and scalability.

\subparagraph{Protocol}
QuKy implements \textit{efficient-BB84} with three polarization states and one decoy level \cite{Ding2017}, matching BB84 security/performance \cite{Tamaki2014} and yielding higher finite‑key rates than two‑decoy schemes \cite{Rusca2018}. Transmitter and receiver fit a 2U 19'' rack for easy integration. Phase randomization via laser gain switching \cite{Kobayashi2014} and a source‑device‑independent QRNG \cite{Avesani2018} (QRN2Qubit) are key components of the system: phase correlations degrade SKR in decoy methods \cite{Tang2013}, while QRNGs avoid PRNG backdoors \cite{cryptoeprint:2016/577}.

\subparagraph{Stability}
QuKy uses the iPOGNAC encoder \cite{Avesani:iPOGNAC}, an all‑fiber, self‑compensating Sagnac‑loop polarization modulator providing calibration‑free, low‑error, long‑term stable operation. A single-phase modulator, asymmetrically placed in the loop, imparts independent CW/CCW phase shifts to synthesize BB84 states on the Poincaré sphere and passively cancel common‑mode drifts. Reported experiments show low QBER over hours in the lab and the field. The hardware can be retimed/rebiased to support time‑bin generation \cite{Scalcon2022}, continuous‑variable modulation \cite{sabatini2025}, and intensity modulation \cite{Berra2023}.

\subparagraph{Synchronization}
\textit{Qubit4Sync} \cite{calderaroFastSimpleQubitBased2020} is a qubit‑aided clock‑recovery algorithm embedding a sparse, known pattern in the quantum stream so the receiver estimates clock phase/frequency from single‑photon timestamps via correlation or ML estimators, then tracks drift adaptively. It is protocol‑agnostic, fiber/FSO compatible, reduces Raman crosstalk and side‑channels by avoiding bright pilots, and has FPGA/SoC real‑time implementations. QuKy uses Qubit4Sync to operate over a single communication channel.

\subsection{Node orchestration and link switching}
As QKD evolves into large-scale quantum networks, a central challenge is the orchestration of QKD nodes.
This involves coordinated management of quantum and classical resources such as link scheduling~\cite{Tayduganov2021}, key routing, load balancing, and fault tolerance. Efficient orchestration maximizes utilization of quantum channels, reduces key distribution latency, sustains continuous key availability for secure applications~\cite{HuguesSalas2019}, and lowers deployment costs by allowing shared use of existing infrastructure thanks to link switching.
The task grows in complexity in heterogeneous networks~\cite{Martin2023}, multi-hop paths, and multiple QKD protocols.
Because orchestration is critical to network-wide secure key delivery, several standards have been proposed~\cite{ETSI_QKD018, ETSI_QKD015, ITU_Y3804, ITU_Y3805}. Most, however, assume a centralized controller, which introduces a potential single point of failure.
The QuKy platform addresses this limitation by extending the ETSI QKD GS 015 interface to support quantum link switching.
Here, a node agent can control a local optical switch, selecting outbound ports toward neighboring nodes.
This allows a single QuKy to be time-multiplexed across multiple QKD links, reducing hardware demands. After initial configuration, key management systems coordinate directly via a proprietary protocol, removing the need for continuous external management.
Crucially, in the case that the central SDQKD controller faults, the cooperating QuKys can keep switching between links and maintain the service operational.
This removes the concern of a series of attacks where the controller is involved, as countermeasures to ignore the central management can be implemented, making it infeasible to take down the network.

Two switching policies have been developed: key balancing, where the resources dedicated to each link are defined with a set of ratios of buffered key to keep, and coordinated switching, where the devices follow a clock dictated by the number of key blocks generated (that is, every $n$ blocks generated by each active link, the switching procedure boots).
The approach allows for future new policies to be added or modified, allowing finer control over the setup.
In the case of optical network maintenance on a fiber link instead, the adjacent nodes skip it after a timeout and try to start the next QKD link in the schedule to keep the service running.
This allows for hardware maintenance without having to reconfigure or interact with the QKD devices.
Moreover, the switching capability greatly reduces the initial cost of introducing the technology on the network, as a new node can be added to the network with just one QKD-capable device and an optical switch, compared to one QKD device per link.

\section{Methods}
\subsection{Analysis of link performances}

The network was released into production in 2025.
To assess the continuous service, we gathered operational use data - comprising Raw and Secret Key Rate (RKR, SKR), QBER, and block-time - for a period of two consecutive months spanning from April to June 2025.
The RKR represents the detection rate at the receiver device, the SKR is the number of secure bits per second that two connected devices deliver, the QBER is the Quantum Bit Error Rate evaluating the imperfection in the qubit transmission, while the block-time 
is the time required to accumulate a block of data (see below).
This gave us a picture of possible long-term trends in the network.
We also batched the data hour by hour to highlight possible daily trends.

\subsection{Description of the network (VenQCI)}
\begin{figure}[h]
    \centering
    \includegraphics[width=1\linewidth]{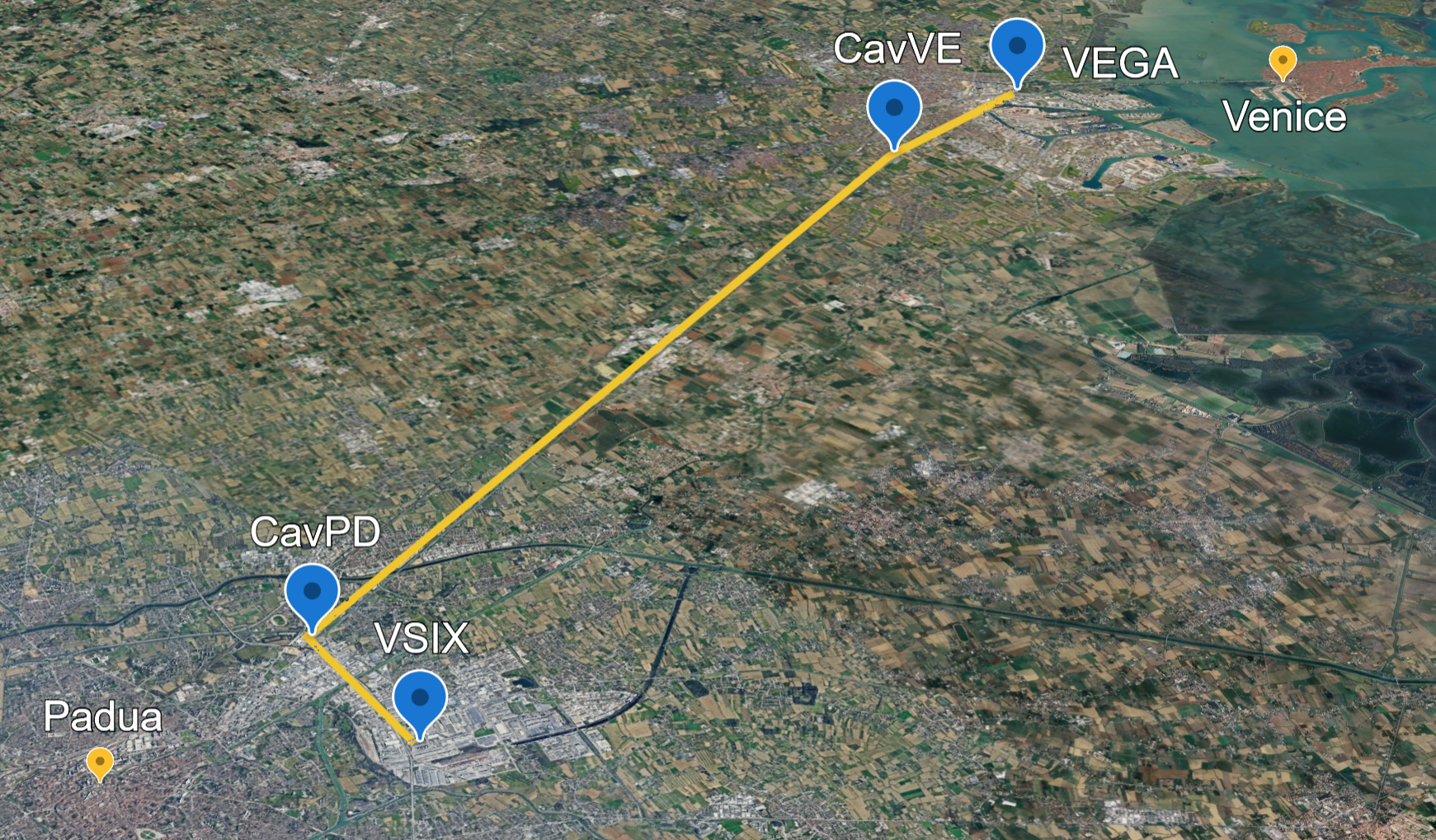}
    \caption{The VenQCI network: VSIX – CavPD (Padova toll booth) – CavVE (Mestre toll booth) – VEGA. Aerial view provided by \href{https://earth.google.com}{earth.google.com}.}
    \label{fig:aerialviewvenqci}
\end{figure}

The network here analyzed consists of four nodes: VSIX - Padova toll booth - Mestre toll booth - VEGA (\vref{fig:aerialviewvenqci}). Despite another fifth node being present, it is currently being used for research purposes and not in an active production environment, which is the focus of this paper. Therefore, we will treat the network as a four-node network in the entirety of the exposition. Each is connected with the following:
\begin{itemize}
\itemsep-0.2em
\item VSIX - Padova toll booth (CavPD): $\sim 5$ km fiber
\item Padova toll booth (CavPD) - Mestre toll booth (CavVE): $\sim 20$ km fiber
\item Mestre toll booth (CavVE) - VEGA: $\sim 5$ km fiber
\end{itemize}

\begin{figure}[h]
    \centering
    \includegraphics[width=1\linewidth]{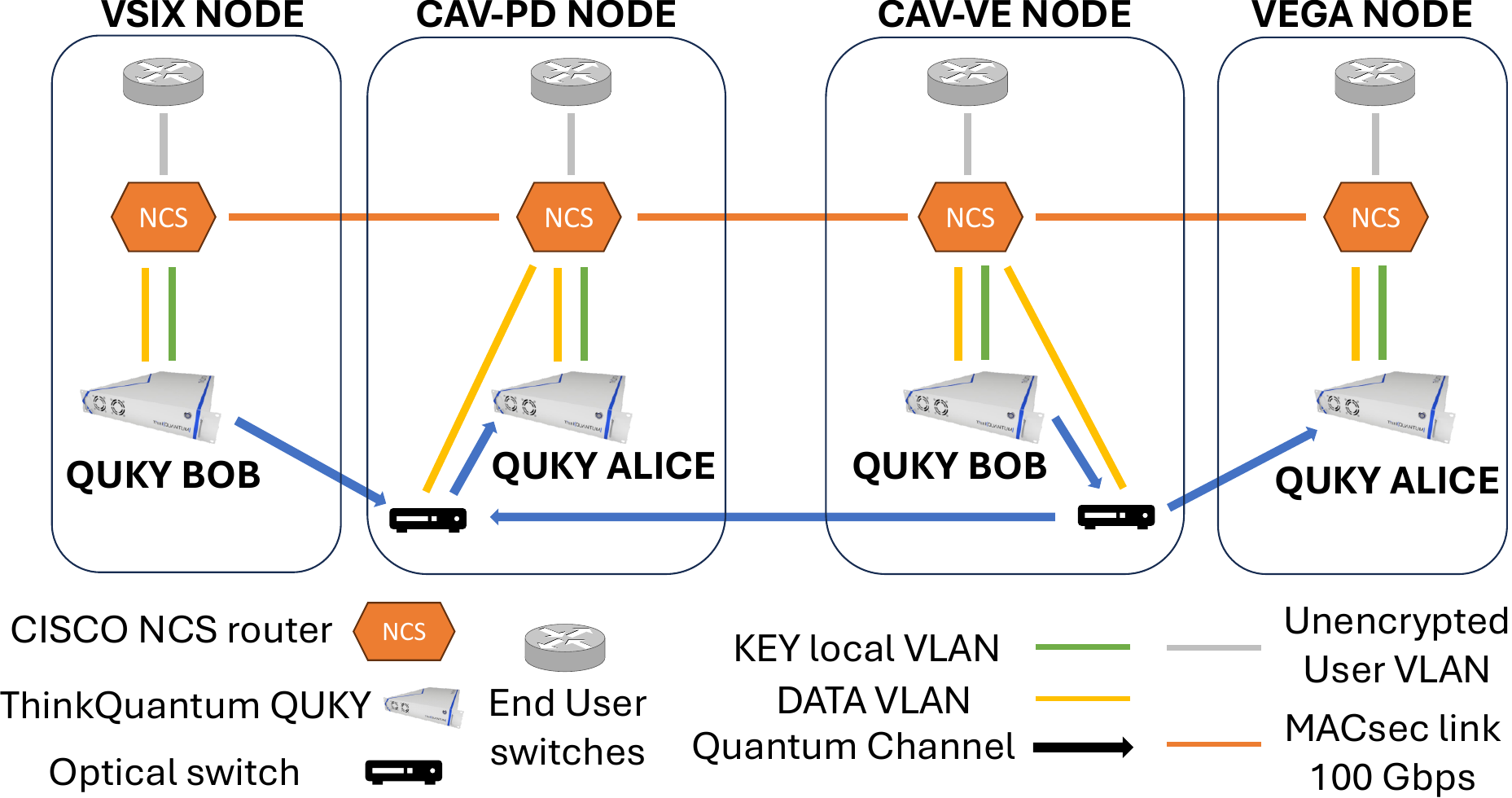}
    \caption{Scheme of the VenQCI network.}
    \label{fig:network}
\end{figure}

The network architecture involves multiple components coordinately operating to enable secure communication, and it's represented schematically in \vref{fig:network}. At the edge of the system, the end users' switches are responsible for transporting unencrypted \textit{VLAN} traffic within the secure perimeter domain.
These switches provide connectivity between nodes through a \textit{Multi-Protocol Label Switching}~\cite{rfc3031} (MPLS) network, ensuring high-performance routing capabilities. MPLS is a routing technique that encapsulates packets from various network protocols and defines the paths between the endpoints through which those packets are transported.
In the backbone of the network are found the \textit{NCS routers}, manufactured by \textit{CISCO}, that implement the MPLS network over encrypted links.
The encryption is provided by \textit{Media Access Control Security}~\cite{MACsec} (MACsec), a layer 2 protocol that ensures data confidentiality, integrity, and authenticity.
The MACsec encryption keys are derived from symmetric keys distributed via the Quantum Key Distribution (QKD) system.
The encrypted MACsec communication is established over a duplex optical link supporting data rates of up to 100 Gbps, utilizing a dedicated single-mode duplex fiber to ensure physical layer security and performance.
The QKD infrastructure is implemented using the QuKy platform developed by ThinkQuantum.
Within this architecture, the quantum channel—used for the transmission of quantum states—is carried over a dedicated simplex optical fiber.
This fiber connects the QKD devices via an optical switch, which is deployed exclusively at intermediate nodes to facilitate dynamic routing of the quantum signal.
In parallel, the classical channel required for QKD protocol execution (such as sifting, error correction, and key reconciliation) is implemented via a dedicated VLAN that is transported over the existing MPLS network.
Once the QKD process has generated symmetric encryption keys, these keys are delivered to the NCS routers using the SKIP~\cite{cisco-skip-01} interface.
This interface operates over a VLAN that is local to each node, ensuring secure and isolated delivery of data to the encryption devices. 

During the booting phase, the MACsec link starts with a pre-shared key until QKD keys are available.
Then, MPLS runs over the new established MACsec link and subsequently carries the QKD's CC data.
New QKD keys are then generated, and the MACsec link refreshes with a new QKD key of 32B each minute.

\section{Results}

\begin{figure}[h]
    \centering
    \includegraphics[width=\linewidth]{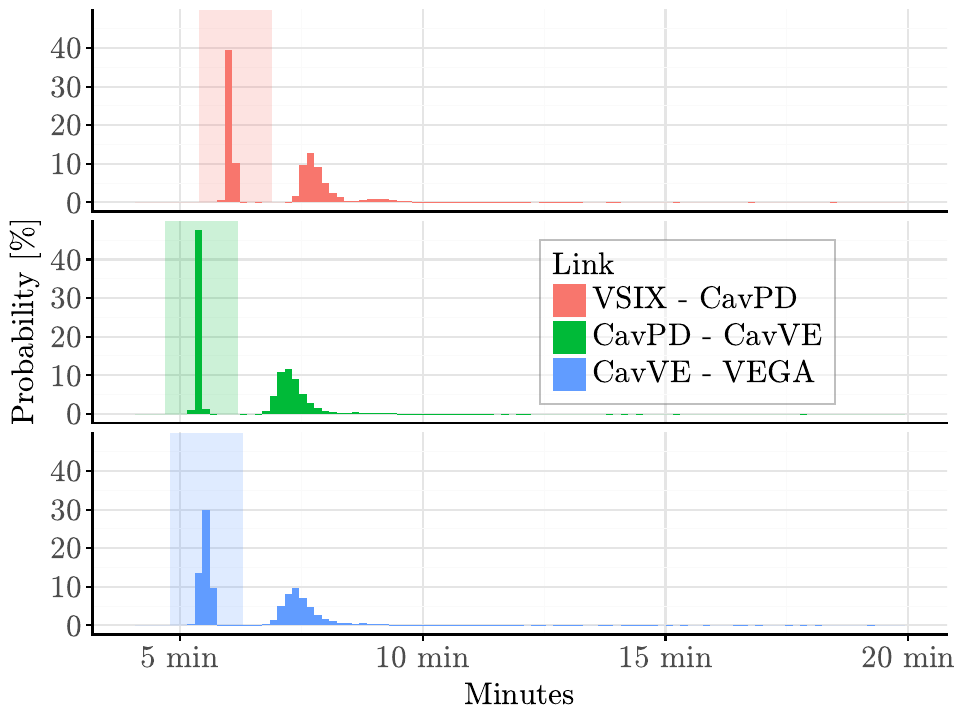}
    \caption{For each link, probability distribution of the time required to accumulate and process 500 kB of sifted key. In each graph are present two peaks, the one on the right corresponding to the first block after the switching (including the initial base alignment time), and the one on the left for the following blocks, which skip this phase (inside the colored area).}
    \label{fig:block-success}
\end{figure}

\begin{figure}[h]
    \centering
    \includegraphics[width=\linewidth]{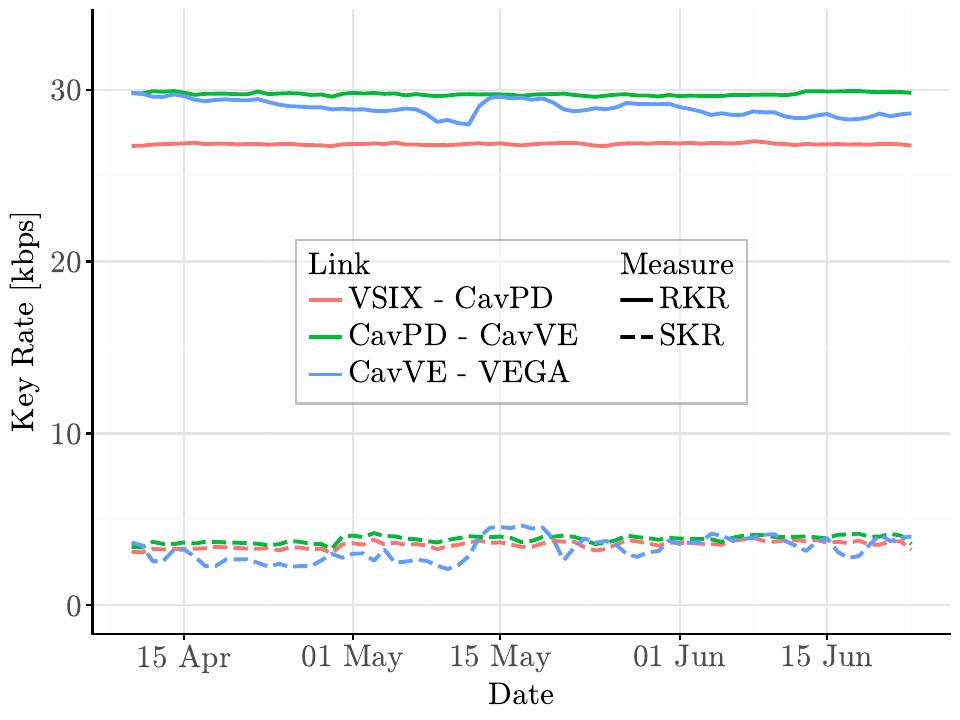}
    \caption{RKR and SKR of each link distinguished by line type. The line represents the mean taken on a daily basis.}
    \label{fig:KR-link}
\end{figure}

\begin{figure}[h]
    \centering
    \includegraphics[width=\linewidth]{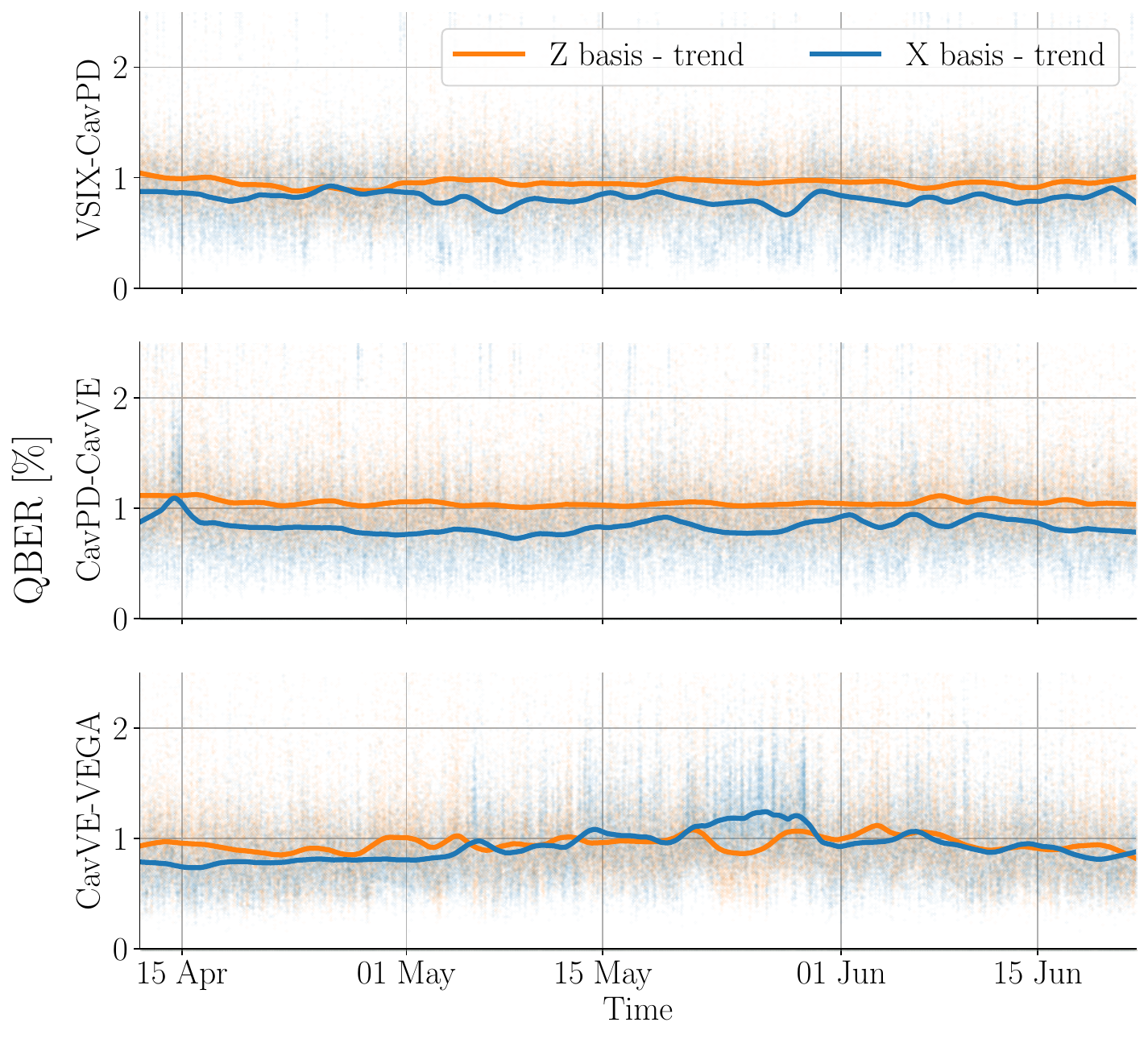}
    \caption{QBER trend for each 2-node link in a two-month period. Experimental data is represented by a scatter plot, while the average is reported as a line plot.}
    \label{fig:qber-link}
\end{figure}

\begin{figure}[h]
    \centering
    \includegraphics[width=\linewidth]{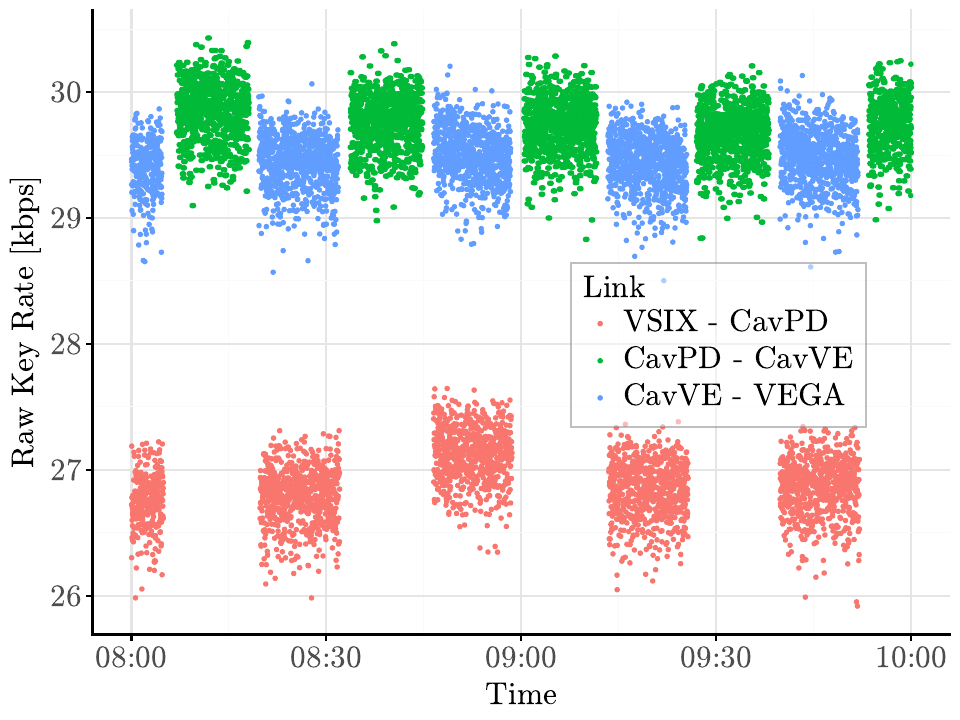}
    \caption{Plot of the switching mechanism sampled from the long-run. The data represents a period of 2 hours on the 13th of May 2025.}
    \label{fig:RKR-link-zoom}
\end{figure}

\begin{figure}[h]
    \centering
    \includegraphics[width=\linewidth]{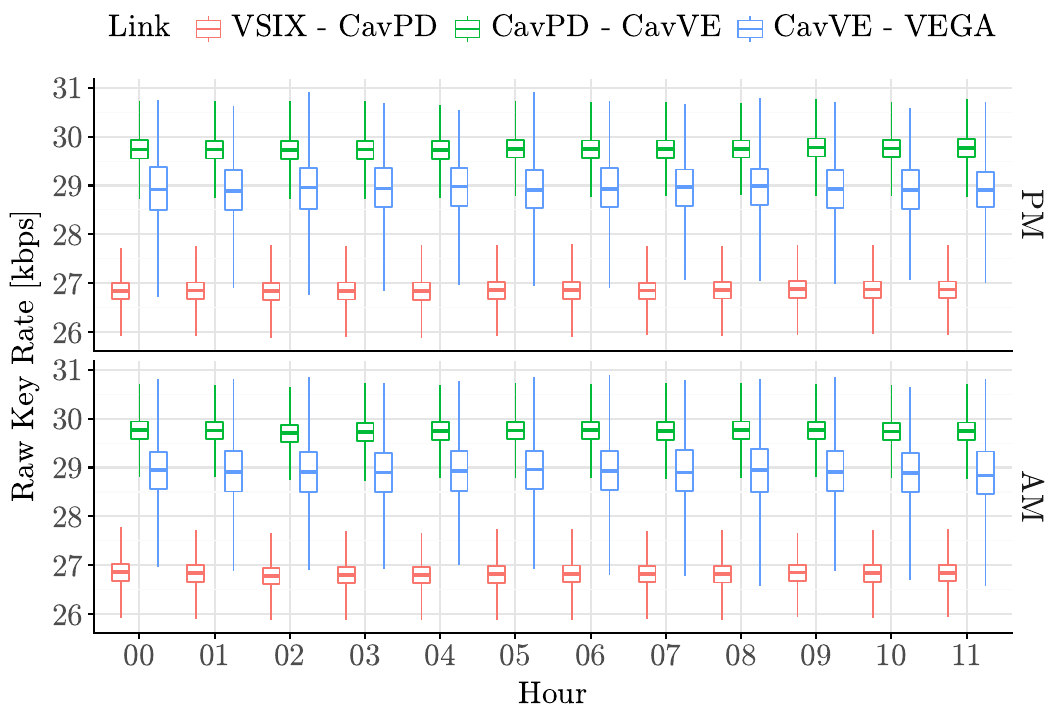}
    \caption{Raw Key Rate by the hour. Box-plot, showing data inside 3 sigmas, where the middle point of the boxes represents the median data point, the box represents the 25\% and the 75\% quantiles.}
    \label{fig:RKR-hour}
\end{figure}

\begin{figure}[h]
    \centering
    \includegraphics[width=\linewidth]{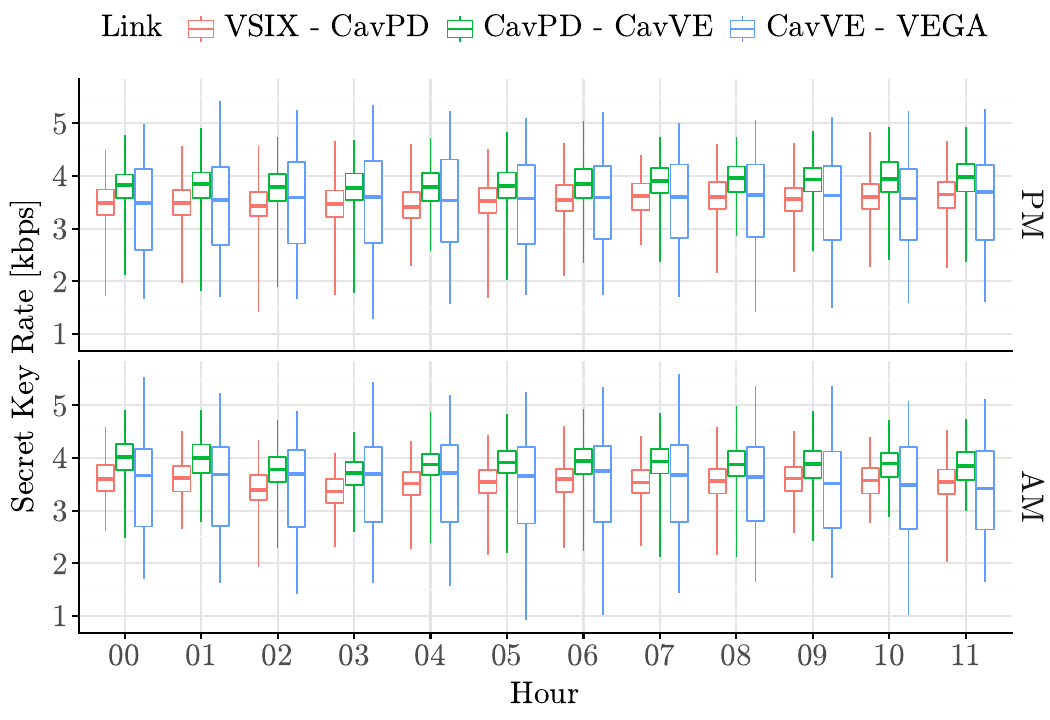}
    \caption{Secret Key Rate by the hour. Box-plot, showing data inside 3 sigmas, where the middle point of the boxes represents the median data point, the box represents the 25\% and the 75\% quantiles.}
    \label{fig:SKR-hour}
\end{figure}

The QKD post-processing is performed starting from batches of 500 kB of sifted detections, which take, on average, approximately 6 minutes to produce.
In \vref{fig:block-success}, the distribution of time necessary to complete and process one of these blocks is shown.
For each link, two distributions emerge; the one on the right corresponds to the first block created after each switch, while the one on the left, in the highlighted area, corresponds to the following blocks.
The former comprises an initial base alignment phase that makes the block take longer with higher variance, while the latter skips this step, leading to more consistent and shorter times.
In \vref{fig:KR-link}, we can see the results for both the RKR and SKR, collected over the two months.
The lines represent the daily average computed for each metric.
The links connecting CavPD to both VSIX and CavVE seem more stable, whereas the one connecting VEGA to the Mestre toll presents a higher variance both in raw and secret key rate.
This, coupled with longer times during data acquisition, points to more significant and less predictable losses that require further investigation in the future.
In \vref{fig:qber-link}, we can see the average and the detailed measurements for the QBER, both in the X and Z bases, with the scatter representing the single data points and the line the 3-day average.
A zoomed-in visualization of the span of two hours, illustrated in \vref{fig:RKR-link-zoom}, shows the switching mechanism at work.
The paired QKD devices collect measurements until they have enough to create two blocks, after which they coordinate to start the other link as pictured.
The \vref{fig:RKR-hour,fig:SKR-hour} show the box-plot of the data collected grouped by hour, over the months.
Here, the middle point of the boxes represents the median data point, the box represents the 25\% and the 75\% quantiles; meanwhile, single events where the key rate dropped outside the main trend by more than 3 sigma were removed, because considered random fluctuations.
This analysis was done to investigate whether some daily trend would appear, such that it would suggest a correlation between link performance and external factors with daily cycles like highway and railway traffic.
Even though there is some fluctuation in the data, we can see how the difference in the median is negligible compared to the interval indicated by the distribution.


After two months, the network has proven to be stable and reliable.
VenQCI is well-positioned to sustain future traffic demands due to its robust key generation capacity, currently exceeding 3 kB/s of SKR on all links.
Given that a single secure connection like those presented in the introduction consumes just 32 bytes of key material per minute, the network can support a substantial number of simultaneous secure links with bandwidth to spare.
This high bit rate not only ensures scalability as the number of users grows, but also provides headroom for more frequent key refresh rates or the integration of additional security services.

The study is limited by the two-month observation window for the full network and by the metropolitan scale of the topology.
Future works should extend the monitoring across seasons and maintenance-intensive periods to further quantify resilience, and investigate scalability toward larger, heterogeneous networks with standardized, interoperable orchestration in alignment with emerging ETSI and ITU-T standards.
Additionally, in a broader view, adapting to key-demand elasticity will require a new dimensioning of switching schedules, buffering, and post-processing throughput.

This defines a clear path for expansion and increased security in communications within the area.
Future plans include the progressive extension of the network and its integration with other segments to form the Italian QCI backbone, enabling a unified, resilient infrastructure at the national scale.
The initially reduced cost, made possible by switching, ensures that the project can follow an incremental development model rather than requiring a single, large-scale deployment.
Meanwhile, multiple sections of the backbone can be developed in parallel and later connected through the inter-domain interface next to be released as ETSI QKD GS 020.
This standard will define a REST-based interface to allow two Key Management Systems (KMS), located in the same trusted node between two network domains, to exchange keys and extend key relay across different implementations and vendors.
This staged approach simplifies the process, minimizes upfront investment risks, accelerates time to operational readiness, and allows for adaptation as technological and operational requirements evolve.
It would  provide a framework for continuous upgrades in performance, security, and interoperability, ensuring the infrastructure remains aligned with broader European quantum communication initiatives.
It can also enhance resilience and flexibility by allowing independent domains to collaborate while maintaining their own security policies.
This encourages the development of vendor-agnostic deployment of multi-domain QKD networks by permitting standardized, secure key exchange between heterogeneous KMS platforms without requiring end-to-end vendor uniformity.

\section{Conclusions}
We have presented the design, deployment, and operational evaluation of VenQCI, an active four-node quantum key distribution (QKD) network that integrates multiple ThinkQuantum s.r.l. QuKy platforms, which exploit efficient-BB84 one-decoy protocol, iPOGNAC-based polarization encoding, and Qubit4Sync synchronization.
The system delivers quantum-derived keys to different infrastructures through ETSI-aligned interfaces and SKIP, enabling periodic MACsec re-keying over a 100~Gbps MPLS backbone.
Our methodology consisted of a two-month multi-node active operation with dynamic link switching in a metropolitan setting.

The field-test results demonstrate the technical viability and stability of the architecture.
The network operation sustained low QBER with elevated raw and secret key generation.
Raw-key batches of 500 kB were formed, on average, approximately every six minutes across links, enabling continuous post-processing and stable secret key rates.
Analyses by the hour did not reveal statistically significant daily trends, suggesting that environmental dynamics and traffic-induced disturbances on the fibers had a limited impact on key production.

From a networking perspective, we introduced and validated a node-agent-driven quantum link switching mechanism that extends the ETSI QKD GS~015 model.
This approach time-multiplexes a single transmitter across multiple neighbors, reducing costs for initial deployment and partially mitigating the risk of a single link sustaining damage.
Moreover, coordinated switching preserved service continuity during fiber maintenance without the reconfiguration of QKD endpoints, evidencing practical resilience in operational environments.

The system integration across layers was validated end-to-end.
QKD-derived keys were consumed for minute-scale 32~B MACsec rekeying, while the MPLS infrastructure dealt with the classical communications once encryption was established.
The bootstrapping sequence derived keys as designed, illustrating a promising path for future quantum networks.

In conclusion, the VenQCI network demonstrates that QKD can be operated as a networked utility rather than a point-to-point solution.
The system yielded stable secret keys suitable for high-throughput infrastructures.
These results provide a practical blueprint for metropolitan-scale quantum-secured services and a foundation for scaling toward larger, interoperable quantum networks.

\section*{Acknowledgements}
A.D.T. acknowledges the financial support of Concessioni Autostradali Venete (CAV) S.p.A. in the framework of the doctoral scholarship agreement 38° Ciclo between CAV and the University of Padova.
This project has received funding from the European Union’s Horizon Europe research and innovation programme under the project "Quantum Secure Networks Partnership" (QSNP, grant agreement No 101114043)” and DIGITAL Simple Grants programme under grant agreement No 101091408. Funding call: DIGITAL-2021-QCI-01.

\section*{Author contributions}
The deployment of the network has been carried out by A.D.T., A.E., M.A., M.V. and L.C.. The data collection, analysis and study has been performed by A.D.T., A.E. and E.B.. This manuscript was drafted and written by A.D.T., E.B. and revised by G.V., L.C. and P.V..

\bibliographystyle{ieeetr}
\bibliography{references}

\end{document}